# Light-stimulable molecules/nanoparticles networks for switchable logical functions and reservoir computing.

Y. Viero,[1] D. Guérin,[1] A. Vladyka,[2] F. Alibart,[1] S. Lenfant,[1]
M. Calame[2] & D. Vuillaume[1,]*

1) Institute of Electronics, Microelectronics and Nanotechnology (IEMN),
CNRS, Univ. of Lille, Av. Poincaré, 59652 cedex, Villeneuve d'Ascq, France.
* Corresponding author : dominique.vuillaume@iemn.fr

2) Department of Physics, University of Basel, Klingelbergstrasse 82, 4056 Basel,
Switzerland.
and
Transport at Nanoscale Interfaces Laboratory, Empa, Swiss Federal Laboratories
for Material Science and Technology, Überlandstrasse 129, 8600 Dübendorf,
Switzerland.

**ABSTRACT**. We report the fabrication and electron transport properties of nanoparticles self-assembled networks (NPSAN) of molecular switches (azobenzene derivatives) interconnected by Au nanoparticles, and we demonstrate optically-driven switchable logical operations associated to the light controlled switching of the molecules. The switching yield is up to 74%. We also demonstrate that these NPSANs are prone for light-stimulable reservoir computing. The complex non-linearity of electron transport and dynamics in these highly connected and recurrent networks of molecular junctions exhibit rich high harmonics generation (HHG) required for reservoir computing (RC) approaches. Logical functions and HHG are controlled by the isomerization of the molecules upon light illumination. These results, without direct analogs in semiconductor devices, open new perspectives to molecular electronics in unconventional computing.

1. Introduction

Networks of molecularly linked metal nanoparticles (hereafter named NPSAN : nanoparticle self-assembled network) represent a powerful approach in molecular electronics to understand fundamental electron transport mechanisms, as well as to study potential applications in electronics and computing circuits. NPSANs with simple molecules (alkyl chains, short π-conjugated oligomers) were used to study metal-insulator transitions,[1] plasmonic,[2] and co-tunneling,[3] for instance. NPSANs were also demonstrated as useful and versatile platforms[4] to study optically-driven molecular switches[5] and redox molecules[6] leading to NPSANs with memory and negative differential resistance behaviors.[7] In the field of electronics and computing circuits, the idea to use a network of self-assembled metallic nano-objets connected by electrically switchable molecules to implement reconfigurable Boolean logic gates though genetic algorithms (the so-called nanocell concept) was originally proposed and simulated by Tour et al.[8] and further explored (simulations) by Sköldberg and Wendin.[9] An experimental demonstration of the training of molecule/nanoparticle networks via genetic algorithms was achieved more recently by Bose and coworkers.[10] However, contrary to previous studies, these authors did not use switchable molecules but the fact that the electronic transport in NPSANs made of alkylthiol capped Au NPs is dominated by Coulomb blockade below 5K (molecules are just linkers between NPs that act as non-linear single electron transistors). In the field of unconventional computing (artificial neural networks), the usefulness of these kinds of networks was also discussed.[11] The groups of Aono and Gimzewski have demonstrated that networks of atomic switches and metallic wires exhibit strongly non-linear electron transport behaviors and complex internal dynamics prone to be used in a reservoir computing (RC) system.[12] RC based on networks of carbon nanotubes dispersed in organic polymers were also proposed.[13]

Here, we demonstrate that NPSANs of Au NPs functionalized by optically-driven molecular switches (azobenzene derivatives) connected in their periphery by 6-terminal Au electrodes exhibit both optically-driven switchable logical operations and switchable strongly non-linear electron transport and dynamical behaviors required for reservoir computing. These behaviors are associated to the light controlled switching of the molecules. We identify 3 main output logical functions with a switching yield of 65%. Remarkably, the devices work at room temperature and fully exploit the functionality of the molecules. Compared to previous devices based on the more generic Coulomb blockade effect in NPs arrays, our NPSAN devices open a richer spectrum of possibilities and a same NPSAN addressed by graphene electrodes instead of gold showed an improved switching yield up to 74%. We also demonstrate how complex non-linearities of electron transport in the NPSANs induce high harmonics generation (HHG) required for RC approaches. We show that HHG can be modified by UV illumination, which paves the way for the processing of multi-input signals through a device that acts as a reservoir computer.

**2. Switchable logical functions.**

The NPSANs consist of a self-assembled network of Au NPs (10 nm in diameter) functionalized by azobenzene-bithiophene-alkylthiol (AzBT) molecules connected by several electrodes (metal or graphene) arranged on a ring (**Figure 1-a** and S1 and S2 in supporting information). More details are given in experimental section and supporting information. This AzBT molecule was chosen owing its high conductance variation upon trans-to-cis isomerization : the cis/trans conductance ratio is up to $7 \times 10^3$ in self-assembled monolayer configuration,[14] and up to 600 in 2-terminal NPSAN.[5d] We have also shown that these molecules embedded in NPSAN can switch reversibly (10 cycles tested in Ref. [5d]).

**2.1. NPSANs with metal electrodes.**

First, the 6-terminal AzBT NPSANs (Figure 1-a) were characterized systematically by measuring the current-voltage (I-V) curves between the 15 possible 2-electrode combinations. The I-Vs were measured for the trans and cis forms of the AzBT molecules (Figure S3-a in supporting information). **Figure 1-b** shows the typical distribution of the switching ratios ($I_{cis}/I_{trans}$ calculated at 2 V from the I-V curves) for 4 NPSANs as the one shown in Figure 1-a where the central ring between the electrodes is 120 nm in diameter. The average on/off ratio is about 8 with a min/max of 2/53. These switching ratios are consistent with our previously reported switching ratios measured on 2-terminal AzBT-NPSANs,[5d] albeit on the low side of the total distribution (between 3 and 620 in that later case).

To demonstrate logic functions, we applied two square voltages on two randomly chosen inputs. These square input voltages were phase-shifted by $\pi/4$ in order to define four Boolean logic input states (**Figure 1**). We simultaneously measured the output currents on the four remaining electrodes. We have chosen periodic signals at low frequency for the sake of simplicity to identify logical functions. We anticipate that any aperiodic combination of signals should work as well, as soon as they have a time scale compatible with the limit of the frequency response of the system (tested in this work up to 20 kHz with AC signals, see section 3). Here, the objective is to demonstrate the switchable functions of the NPSAN rather than to establish a record frequency limit. We have recently demonstrated that tiny molecular junctions (made on 10 nm Au nanodot electrodes) can work up to 17 GHz,[15] it is likely that NPSANs can work at frequencies higher than 20 kHz, the exact frequency limit being set by the parasitic capacitances of the electrodes, which were not optimized for high frequencies in this work as it was not the main focus of the study.

**Figure 1-c to f** shows the typical results for one configuration with input voltages $U_1$ and $U_2$ (amplitude 5 V) applied on electrodes 4 and 2, respectively. The output currents are shown for the AzBT in trans (red traces) and cis (blue traces) states. Several behaviors are observed. On electrode 3 (Figure 1-d), for the trans state, the output current is a copy of the input voltage $U_2$ applied on electrode 2, $I_{3trans}=g_{32trans}U_2$ with $g_{32trans}$ the trans-AzBT-NPSAN conductance between electrodes 2 and 3. We call this function "pass" in the following. When the AzBt are in their cis state, we observed an increase in the current, as expected since we have previously demonstrated[5d, 14] that cis-AzBT-NPSANs are more conductive than trans-AzBT-NPSANs (see also Figure S3-a in supporting information). More important, the shape is modified, now the output current is a weighted sum of the two input voltages $U_1$ and $U_2$ : $I_{3cis}=g_{34cis}U_1 + g_{32cis}U_2$. The conductance values are $g_{32trans}$ = 2 pS, $g_{32cis}$ = 8 pS and $g_{34cis}$ = 18 pS. The increase of $g_{32}$ by a factor 4 is in agreement with the statistics measured for standard I-V curves on these 6-terminal NPSANs (Figure 1-b). On electrode 5 (Figure 1-e), the situation is simpler. We did not measure a current in trans, which implies that $g_{52trans}$ and $g_{54trans}$ are below the sensitivity of our equipment (a current sensitivity of 100 fA, or 0.02 pS at 5V), we call this case "noise". In the cis form, we observed a "pass" function, $I_5 = g_{52cis}U_2$ with $g_{52cis}$ = 8 pS. On electrode 6 (Figure 1-f), we have a "sum" function for the trans state, $I_6 = g_{64trans}U_1 + g_{62trans}U_2$ with $g_{64trans}$ = 0.4 pS and $g_{62trans}$ = 2 pS. After switching to the cis isomer, we have a "pass" function, $I_{6cis}=g_{64cis}U1$ ($g_{64cis}$ = 18 pS). On electrode 1 (Figure 1-c), we again observed "noise" for the trans state and a "sum" function in the cis state with $I_{1cis} = g_{14cis}U_1 + g_{12cis}U_2$ ($g_{14cis}$ = 2 pS and $g_{12cis}$ = 14 pS). We note that $g_{14cis}$ is the lowest conductance measured in this NPSAN in the *cis* case, this may be due to the fact that electrodes 1 and 4 are separated by the largest gap corresponding to the full diameter of the electrode ring, ca. 120 nm in that case, defined by the 6 electrodes. To summarize, we observed 3 output functions "noise, "pass" and "sum", and 4 types of reconfigurations upon trans to cis isomerization : "pass" to "sum", "noise" to "sum, "noise" to "pass" and "sum" to "pass".

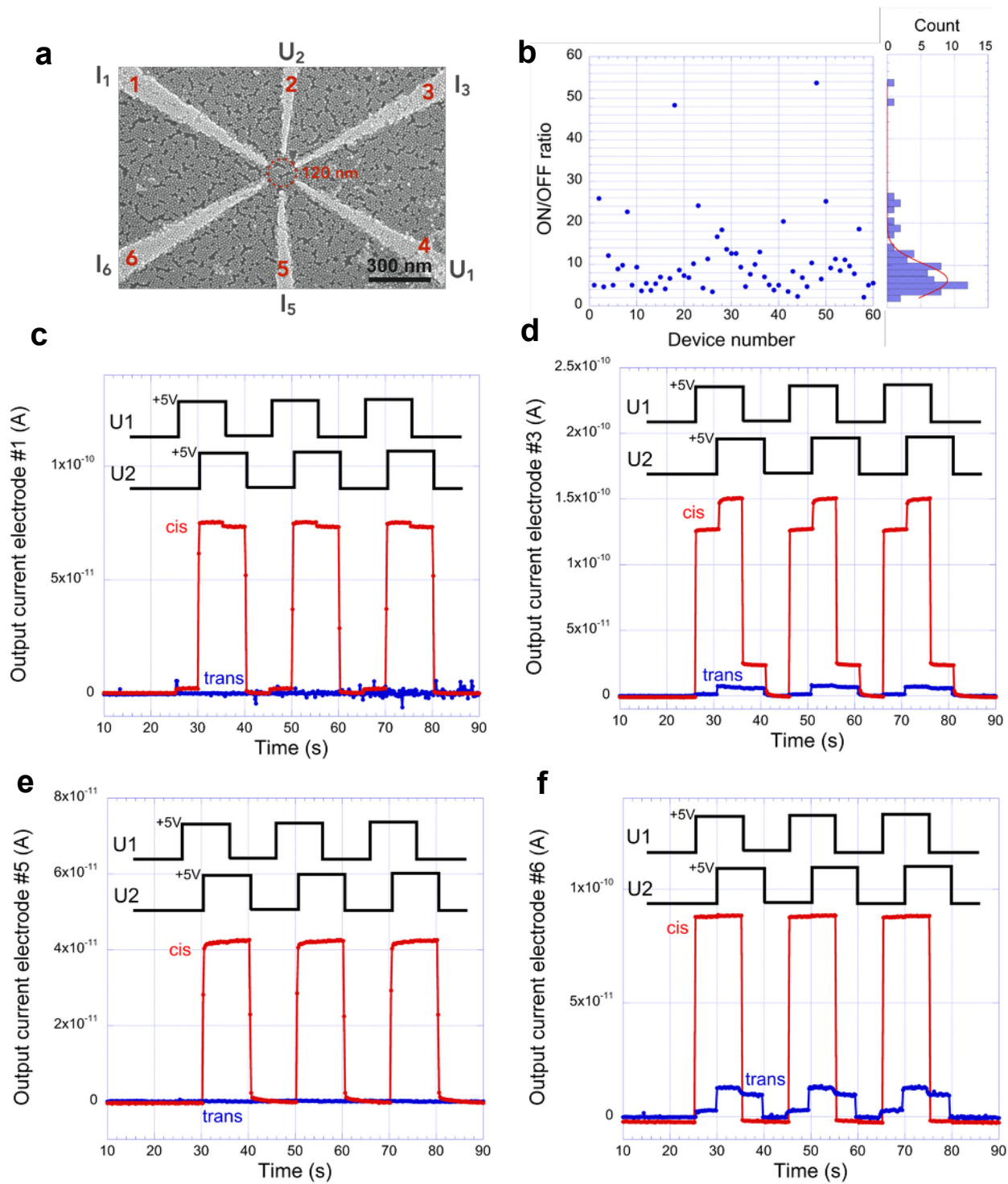

**Figure 1. (a)** SEM image of the NPSAN deposited on 6-electrode pattern (central ring is 120 nm in diameter). **(b)** On/off ratios, $I_{cis}/I_{trans}$, calculated at 2 V form the 60 I-V curves (see Figure S3-a in supporting information) measured from combinations of pair of electrodes in 4 NPSANs when the AzBT molecules are in the trans isomer or cis isomer. **(c-f)** Output currents simultaneously measured on electrodes 1, 3, 5 and 6, respectively, when the AzBT molecules are in the trans isomer (blue lines) and cis isomer (red lines).

We repeated this experiment on a given NPSAN choosing the 2 input electrodes arbitrarily and on 15 different NPSANs. Out of a total 200 outputs measured in both trans and cis isomers, we derived statistics for the trans and cis states (Table 1) and for the type of reconfiguration upon trans to cis isomerization (Table 2). For the trans-AzBT-NPSAN, the most frequent cases are "noise" and "pass" (no signal or only one input transmitted though the NPSAN), while the "pass" and "sum" dominate the behavior of the cis-AzBT-NPSAN. A switching of the NPSANs is observed in 65.5% of the measurements, among them the "pass" to "sum" switching is the most frequent (28.5 %).

| Type of function | Occurence AzBT in *trans* | Occurence AzBT in *cis* |
|---|---|---|
| Noise | 37 % | 2 % |
| Pass | 58 % | 50 % |
| Sum | 5 % | 48 % |

**Table 1.** Statistic of the different functions observed in the NPSANS with the AzBT molecules in the trans and cis isomers.

| Type of event | Occurence (metal electrodes) | Occurence (graphene electrodes) |
|---|---|---|
| noise to pass | 19 % | 25% |
| noise to sum | 16 % | 21% |
| pass to sum | 28.5 % | 25% |
| sum to pass | 2 % | 3% |
| switching | 65.5 % | 74% |
| no-switching | 34.5 % | 26% |

**Table 2.** Statistic of the switching events for NPSANs with metal and graphene electrodes.

**2.2. NPSANs with graphene electrodes.**

Graphene represents a potentially interesting contact electrode material for hybrid molecular systems thanks, in particular, to its high structural stability.[16] The 2D nature of graphene also provides a smoother contact interface for NPSAN devices compared to Au electrodes. Overall we measured 44 NPSAN devices with graphene electrodes. As before, the device resistance was extracted at a bias voltage of 2V (see Fig S3-b in supporting information for typical I-V characteristics). **Figure 2-a** shows distributions of the resistances before (trans-AzBT) and after (cis-AzBT) UV light illumination. **Figure 2-b** shows the distribution of on/off ratios for all measured devices.

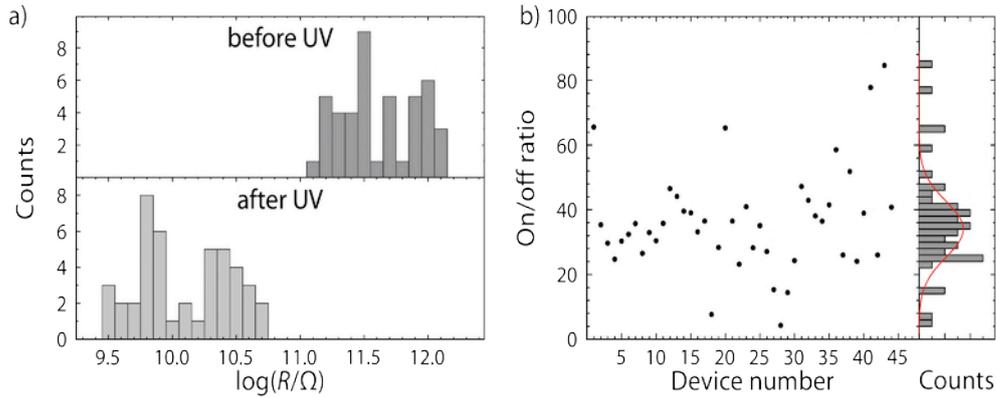

**Figure 2.** Electrical characterization of graphene NPSAN devices. **(a)** Distribution of resistances extracted at V=2V for different devices before (trans state) and after UV illumination (cis state). See typical I-Vs Figure S3-b in supporting information. **(b)** On/off ratio for all measured devices.

The resistances are distributed over a decade with the AzBT in both trans and cis isomers. The off/on ratio (calculated as $R_{trans}/R_{cis}$) is centered at about 35 (with min/max at 4/84). Compared to NPSANs with metal electrodes, we observed that the dispersions of the device resistances is reduced with graphene electrodes (around a decade, Figure 2-a and 2-b) compared to Ti/Au electrodes (around two decades, see Ref. [5d]). This is attributed to a better electrical contact at the interface between graphene and NPSAN. The average on/off ratio is improved in the graphene electrode case (∼ 35 - Figure 2-b, instead of ∼8, Figure 1-b), which can be explained by a reduction of the contact resistance at the NPSAN/graphene interface. A large parasitic contact resistance tends to decrease the measured on/off ratio with respect to the intrinsic on/off ratio of the NPSAN itself. We attribute these improved performances to the reduction of electrode thickness when using graphene (less than 1nm compared to ~11 nm for Au electrodes)[17], the AzBT/NP network being less deformed/distorted at the electrode edges. The min/max values of the distribution of the on/off ratio are similar in both cases (Au and graphene), this dispersion reflects the molecules and NPs organisation inside the NPSAN, less affected by the quality of the contacts at the electrodes. The switching of the logical functions is also improved up to 74% (see Table 2, 64 outputs recorded). Graphene electrodes are thus a valuable solution to improve the global performances of switchable logic functions of the NPSANs.

**3. Switchable high harmonic generation (HHG).**

We now examine the behavior of the NPSANs when we apply sinusoidal signals at two frequencies (**Figure 3-a**). We performed these experiments only for the NPSAN with Au electrodes. This is justified since we aim at studying the complex non-linearities and dynamics electron transport occurring inside the AzBT and NP network itself, in relationship with the optically switchable states of the AzBT molecules, rather than to improve device performances. We have shown (section 2) that graphene electrodes mainly reduce the contact resistance without modifying the general behavior of the NPSANs. The same effect is likely to occur for HHG, i.e. a global increase of the electrical conductance of the NPSANs. Here, we applied two signals at different frequencies to highlight simultaneously the generation of harmonics for a given input signal, but also more complex phenomena such as intermodulation distortion due to interactions between two input signals (see below). Due to the non linearities in the I-Vs, the output currents are not simple combinations of the inputs (see Figure S4 in supporting information) and the spectrum analyzer reveals the generation of high harmonics (**Figure 3-b to e**, Figure S4-c in supporting information), i.e. integer multiple of the frequency of the applied signal, for both input signals (up to the 10th harmonic in some cases, see Figure S4-c in supporting information). For each input signal, the total harmonic distortion (THD) is given by:

$$THD = \frac{\sqrt{\sum_{n=2}^{\infty} C_n^2}}{C_1} \qquad (1)$$

where $C_1$ is the amplitude of the fundamental signal and $C_n$ ($n \geq 2$) the amplitude of the harmonics (THD values are directly reported in Figures 3-b to e for the two input signals). Note that some of the peaks (labeled "x") are not clearly identified (i.e. they do not correspond to the expected frequencies of the harmonics of signals A or B, this will be discussed later). A reference experiment replacing the NPSAN by a network of precision resistors showed (Figure S4-d in supporting information) no harmonic generation. Thus, the HHG is not induced by our measurement equipment, but can be ascribed to the AzBT-NPSANs.

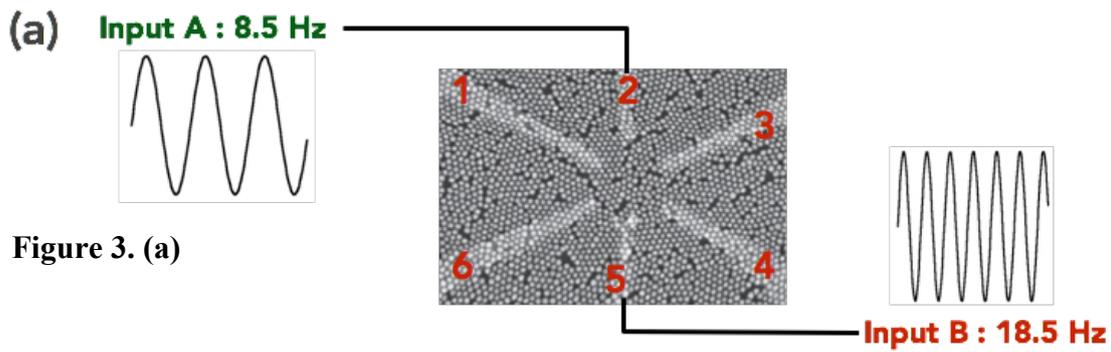

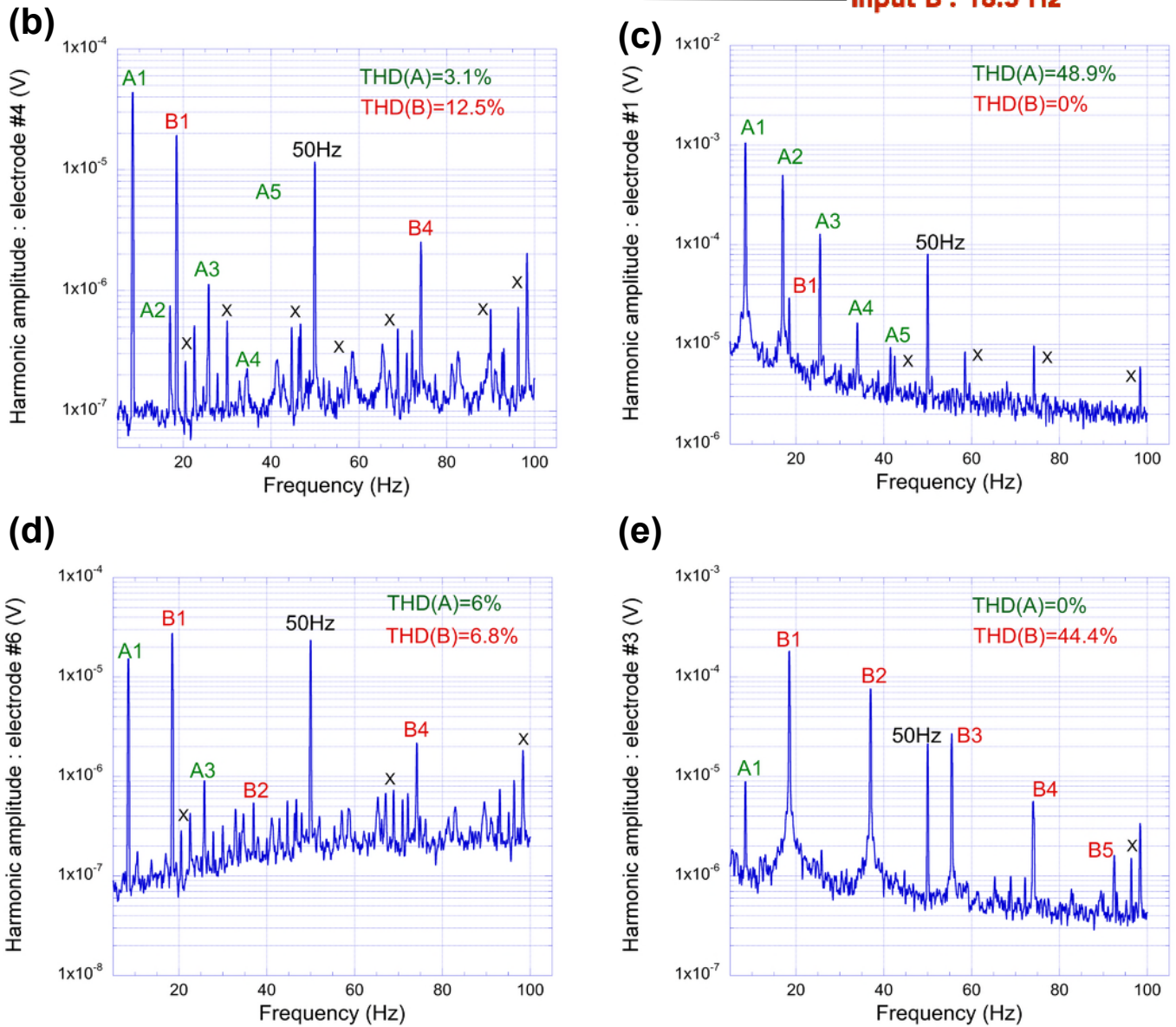

**Figure 3. (a)** Typical configuration for HHG experiments. Two sinusoidal signals, signal A at 8.5 Hz and signal B at 18.5 Hz (peak-to-peak amplitude VPP=8V for both) are applied at electrodes 2 and 5, respectively. **(b-e)** The 4 output currents are measured by the dynamic signal analyzer (FFT) and the THD are calculated for both signals A and B. HHG peaks are given in volt, the output

currents are amplified by a current-voltage amplifier before FFT analysis. The HHG peaks are labeled as Ai (i=1 for the fundamental, i=2 for the 2nd harmonic, etc…) and Bi for harmonics corresponding to the A and B input signals, respectively. The AzBT molecules are in the trans isomer. Some peaks are not identified (labeled x) and are discussed later (see text).

**Figure 3** shows that different behaviors of the 4 outputs are observed for trans-AzBT-NPSAN. We clearly observed heterogenous responses of the 4 outputs, the THD for A and B signals varying from almost 0 to about 50%. For a given output, the THD values are stable when we vary the signal input signal frequencies in the range 1 Hz to 2 kHz. The same heterogeneous situation is observed after the cis isomerization of the AzBT molecules but with a global increase of the HHG (higher THD) as shown in **Figure 4-a and b**. We clearly observe that HHG is more efficient for the cis-AzBT-NPSAN with an increase of the THD for both input signals. We measured 7 NPSANs, a statistical analysis (**Figure 4-c**) confirms that THD is globally higher for the cis-AzBT-NPSAN. The average THD is $\simeq$ 19% in the trans state and $\simeq$ 25% in the cis state. We have observed similar HHG behavior with input signals up to 20 kHz.

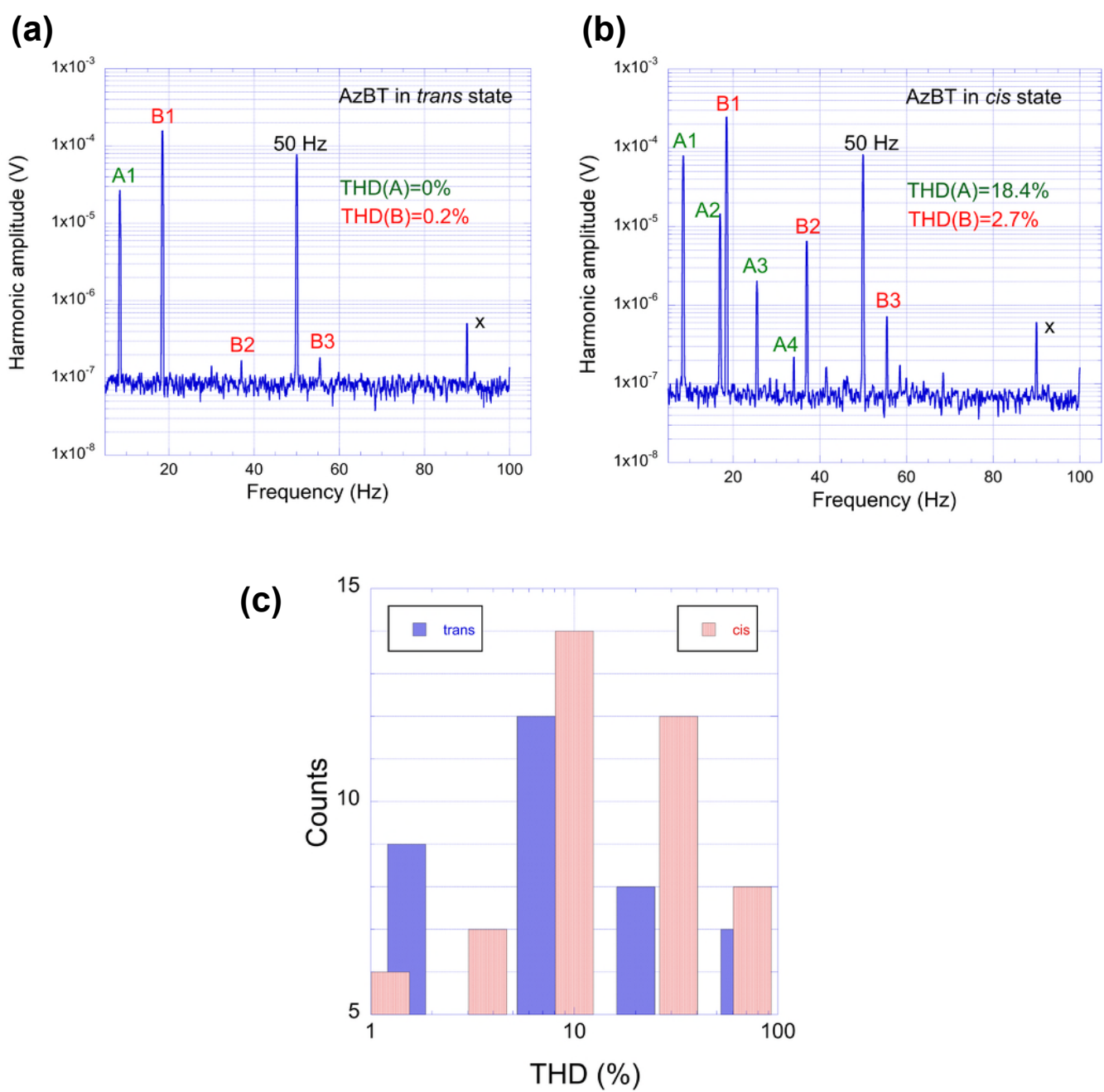

**Figure 4.** Signal frequency spectra for the same NPSAN output when **(a)** the AzBT molecules are in the trans state and **(b)** in the cis state showing a clear increase of the THD related to both signals A and B. **(c)** Histogram of the THD related to signals A and B and calculated from 7 NPSAN samples and for the two isomers of the AzBT molecules. The HHG peaks are labeled as Ai (i=1 for the fundamental, i=2 for the 2nd harmonic, etc…) and Bi for harmonics corresponding to the A and B input signals, respectively.

In some cases, the THD is not significantly modified by the AzBT isomerization but a more complex behavior is observed with the appearance of intermodulation distortion (IMD) in the cis-AzBT-NPSAN (**Figure 5**) up to the 7th-order distortion product. The intermodulation distortion product O is defined by:

$$O = |a_1| + |a_2|$$
$$f_n = a_1 f_1 + a_2 f_2 \qquad (2)$$

where $f_n$ is the frequency of the observed IMD peaks, $a_1$ and $a_2$ integers (>0 or <0), $f_1$ and $f_2$ the fundamental frequencies of the input signals (here 8.5 and 18.5 Hz). Figure 5 shows IMP intermodulation product varying from O=2 (e.g. peaks B1-A1, B1+A1) to 6-7 (O=6 for peak B5-A1 where B5 is the 5th harmonic of signal B, and O=7 for peak labeled B5-A2 where A2 is the second harmonic of signal A). Note that O is usually defined as the sum of existing frequency combinations and does no take into account their weight (i.e. amplitude). To avoid the contribution of noise, we have taken into account only frequencies with a weight greater than 5% of the fundamental. In addition to HHG peaks related to each signal A and B, and IMD, we also observed interharmonics, i.e. non-integer multiples of the input signals (e.g. see peaks labeled B1.5, B2.5 and B3.5 located at 1.5, 2.5 and 3.5 time the fundamental frequency of the input signal B - Figure 5-b). Albeit IMD was also observed in few cases (ca. 14% of the cases) for the trans-AzBT-NPSAN, which may explain the additional peaks (labeled as "x") in the spectra shown in Figure 3, IMD is more frequent (ca. 50% of the cases) for the cis-AzBT-NPSAN (Figure 5-b).

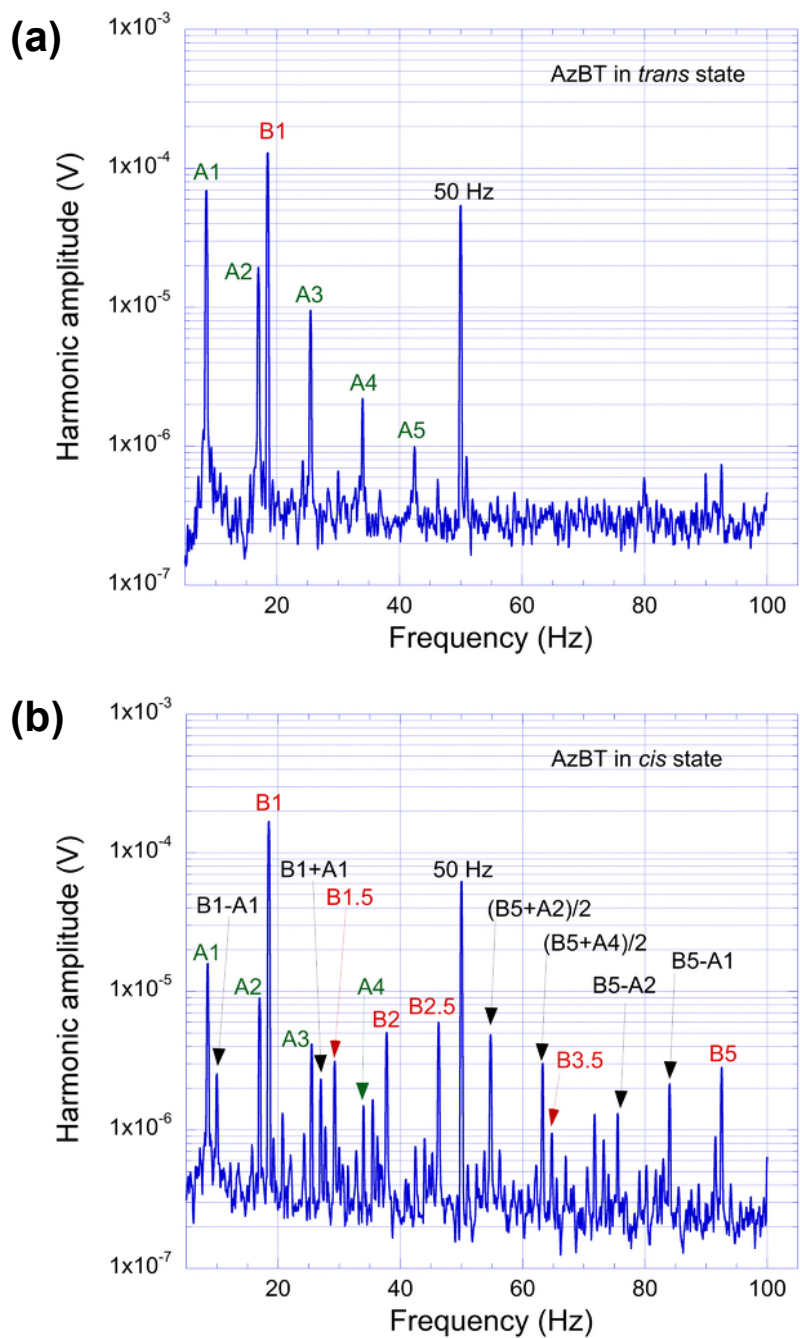

**Figure 5.** Signal frequency spectra for the same NPSAN output when **(a)** the AzBT molecules are in the trans state and **(b)** in the cis state showing IMD and interharmonics. The HHG peaks are labeled as $A_i$ (i=1 for the fundamental, i=2 for the 2nd harmonic, etc…) and $B_i$ for harmonics corresponding to the A and B input signals, respectively. Intermodulation distorsion (IMD) peaks are labeled by a combination of $A_i$ and $B_i$. Interharmonics are labeled with non-integer i values.

## 4. Discussion.

### 4.1. Switchable functions.

The switchable functions called "pass" and "sum" (see Table 1 and 2) can be translated in Boolean logic gates. The "pass" function is simply a follower and the "sum" corresponds to the "OR" or "AND" logic gate, depending on the threshold of the output signal chosen to distinguish the logic state "0" and "1". For a high threshold, in the cis case (see Figure S6 in supporting information), the "sum" function corresponds to the "AND" logic gate (output = "1" if both inputs are set to "1"), while we have the "OR" logic gate (output = "1" if one or the two inputs = "1") for a low level threshold.

Compared to the nanocell approach,[8-9] the advantage here is that we do not need a programmable sequence applying specific signals on the electrodes to switch the molecules in the required states, nor a sophisticated genetic algorithm.[8, 10] Contrary to previous results,[10] our NPSAN devices work at room temperature (not 5K) and fully exploit the functionality of the molecules (and not Coulomb blockade in alkylthiol capped Au NPs). Here, the switchable functions are due to the conductance switching of the molecules associated to their trans-cis isomerization. A control experiment with alkylthiol functionalized Au NPs did not show any significant conductance switching in the NPSAN,[5d] nor logic function reconfiguration. We note that we have not observed other logic functions such as XOR (exclusive OR) or XNOR (the complement of the XOR function). An alternative to implement XOR or XNOR with NPSAN devices would be to have NDR (negative differential resistance) I-V characteristics[18] (that we did not observed in our system with AzBT molecules) or to have some resistive switching induced by the input signal (i.e. voltage dependent switching).[19] Achieving these features in NPSAN remains an open challenge. This may be possible with NPSAN made of other molecules, e.g. we have observed NDR in NPSANs with redox molecules.[7b] Another possibility is to use a neural network approach using these NPSANs (see discussion below).

The variability of the observed functions at the several output electrodes (Figure 1) is likely to depend on the topology (and its modification) of the molecular conducting pathways into the NPSANs upon trans to cis isomerization. The exact number and precise location of AzBT molecules undergoing a trans-to-cis isomerization in the network is difficult to gauge and percolation effects can be expected.[4] A direct and experimental observation of the molecular topology reconfiguration within the network is not straightforward. An experimental visualization of these molecular conducting pathways by conducting-AFM, for example, would require a better resolution than available with our C-AFM equipment.[20]

The foreseen possible applications are in the field of organic opto-electronics. The first advantage of the NPSANs is compactness. Here, a logic gate occupy an area of ∼$10^{-2}$ μm$^2$ (central ring of the NPSAN, excluding wires and connecting pads) while it takes ∼100 μm$^2$ or more (also excluding wires and connecting pads) for Boolean logic gate using OFET.[21] The second advantage is the simplicity to reconfigure by light illumination the implemented logic gates. In principle, in standard organic electronics, it should be possible to implement optically-driven logic gates using more sophisticated optically switchable organic transistors already demonstrated at the single device level,[22] but not yet integrated in circuits and logic gates. Even in that later case, the compactness advantage of the NPSANs holds.

**4.2. Reservoir computing.**

If the great variability of the output signals may be a drawback for classical computing (based on Boolean gates and von Neumann architecture), it is a mandatory condition for neuro-inspired computing and especially for reservoir computing (RC). In this later case, the reservoir is a complex network of highly interconnected non-linear elements, which exhibits non-trivial evolution though dynamics in the network.[23] The non-linear dynamics in the network can be useful for implementing RC.[23c] We investigated how complex non-linearity in the AzBT-NPSAN leads to high harmonics generation (HHG), one of the prerequisites of reservoir computing (RC) approaches.[23a, 23b] Simulations have shown that HHG in a network of atomic switches connected by metallic wires can be used in RC.[12c]

Again, the heterogeneous HHG responses (Figure 3) depend on the topology of the conducting pathways into the NPSANs and the degree of non-linearity in the NPSANs. The weaker HHG in trans AzBT NPSANs is clearly related with the observation that a majority of the I-V curves measured on the 15 combinations of pair of electrodes in a given NPSAN (Figure S7 a and b in supporting information) are almost linear, except few of them (3 out of 15), these later corresponding to the situation where we have observed the highest HHG. The THD increase in the cis AzBT-NPSANs (Figure 4) is clearly related to a stronger non-linearity of the I-V curves as shown in Figure S7-c and d in supporting information. For the cis AzBT NPSAN, all the 15 I-V curves show a strong non-linear behavior. The fact that I-V curves are more linear when the AzBT molecules are in the trans form may be related to the fact that the conductance though the molecules between two adjacent NPs is lower. The electron transfer in the NP-molecules-NP junction is mainly dominated by tunneling and it is known that the linear part of the tunneling I-V is valid as long as the voltage across a NP-molecules-NP junction is below $\sim(\Delta/3e)$,[24] with e the electron charge and $\Delta$ the energy barrier at the metal/molecule interface, i.e the energy difference between the metal Fermi energy and the energy of the molecular orbital involved in the charge transport. The voltage across a NP-molecules-NP junction is roughly approximated by the applied voltage between the contacting electrodes (e.g. 4 V in Figs S7) divided by the number of such junctions in series (~ 10-15 for the NPSAN with a diameter of 120 nm, Fig. 1-a). We have previously shown that this energy barrier is higher for the trans AzBT (2.3 eV) than for the cis AzBT (1.9 eV).[14] In other words, with a lower energy barrier height, the currents for the cis AzBT start to deviate from linearity at lower applied voltages (as shown in Fig. S7, supplementary information), resulting in a stronger HHG.

Similarly, the IMD is more frequent (occurrence of ca. 50%) when the AzBT molecules are in the cis isomer (Figure 5). In few cases (ca.14%), IMD was also observed with molecules in their trans state. Moreover, in that later case, the intermodulation product is lower (e.g. O = 4, Figure S8 in supporting information) than for the cis case for which O up to 7 was observed (Figure 5-b). IMD occurs at frequencies that are the sum and/or the difference of integer multiples of the fundamental frequencies. In our experiments, we also observed interharmonics (Figure 5), i.e. the presence of spectral components at frequencies that are not integer multiples of the fundamental frequencies.[25] Interharmonics can be caused by oscillations occurring, for example, in systems comprising series of parallel capacitors. This is consistent with the fact that the molecular junctions between adjacent NPs in the network have also a capacitance contribution. The entire networks can be modeled by a network of resistance-capacitance elements connected to each other by the Au NPs.[26] One of the fundamental properties of a reservoir is to project an input space into the highest possible dimensional output space (called the feature space). In this work, we investigated this projection by a focus on HHG, interharmonics and IMD induced by the NPSAN devices. We reveal that we were able to increase the nonlinearities and consequently the feature space by the optical switching of our molecular linkers.

By analogy with artificial neural networks approaches, the NPSAN device can be seen as a single layer perceptron (with multiple input and multiple output) where each input/output terminals are weighted by the NP/molecular pathways configuration. Note that the demonstration of switchable logical functions only takes advantage of static characteristics of the NPSAN devices. So far, single layer structure was not able to implement XOR or XNOR functions (and not observed as reported in section2) while a two layer neural network can elegantly solve this problem. If we consider now the concept of reservoir for these NPSAN devices, such a XOR or XNOR function could be available thanks to the addition of a read out layer (i.e. the system is then a two layers structure). Further work that we will consider will explore more complex computing functions based on RC approaches but remain out of the scope of the present paper.

## 5. Conclusion

In conclusion, we demonstrate that rich and complex behaviors emerge from optically switchable molecules connected by metal nanoparticles. HHG in these NPSANs demonstrate that they exhibit strongly non-linear and internal dynamic behaviors, making these systems promising for reservoir computing. The use of graphene electrodes for NPSANs provide smoother contact edges and result in improved performances of NPSANs as switchable logical functions. A further implementation of NPSAN in a practical implementation of a reservoir computing should benefit from the global improvement of NPSAN performances with graphene electrodes. In addition, the use of light-switchable AzBT molecules in these networks open the door for light-stimulable RC, as an exemple of Evolution-in-Materio concept.[27]

**6. Experimental section**

*Device Fabrication. NPSANs with metal electrodes.* Coplanar 6-terminal nanogap metal electrodes were fabricated using standard electronic lithography processes. We used a silicon wafer covered by a 200 nm thick silicon dioxide. Ti (1 nm thick)/Au (10 nm thick) was deposited and patterned by e-beam evaporation and lift-off. The coplanar electrodes are arranged around a circle with a diameter comprised between 80 to 120 nm (Figure S1 and more detail in supporting information).

*Device Fabrication. NPSANs with graphene electrodes.* Graphene for the electrodes was grown using chemical vapor deposition (CVD) on copper foil at 1000°C using methane as a precursor and then transferred on Si/SiO2 substrate using wet transfer method.[28] Graphene electrodes were fabricated by reactive ion etching using PMMA as an etching mask. Individual device consists of 8 graphene electrodes (see supporting information, Figure S2-a) with a ring size between the electrodes of about 0.5–2 μm (Figure S2-b).

*Preparation and Deposition of the AzBT-AuNPs Monolayers.* The synthesis of the azobenze-bithiophene-alkylthiol (AzBT) molecules is described elsewhere.[29] For the functionnalization of the Au NPs (10 nm in diameter) and the deposition of a monolayer of AzBT-NPs on the electrodes, we followed the same process as in our previous work.[5d] In brief, citrate or oleylamine capped Au NPs were synthesized according to known recipes,[30] and a protocole of ligand exchange was used to functionalize the Au NPS with AzBT molecules.[31] The effectiveness of the synthesis and AzBt functionalization was assessed by UV-vis absorption and XPS experiments,[5d] leading to a density of ca. 2560 AzBT molecules per NP (or 3.2 molecules/nm$^2$ - see details in Ref. [5d]). Then, a monolayer of AzBT-NPs was formed at the water surface and transferred to the patterned substrate by the Langmuir method following the process described in Ref. [32] (Figures S1-b and S2-c in supporting information).

*Electrical Measurements.* The electrodes were contacted with a micromanipulator probe station (Suss Microtec PM-5) placed inside a glovebox (MBRAUN) with a strictly controlled nitrogen ambient (less than 0.1 ppm of water vapor and oxygen). Such a dry and clean atmosphere is required to avoid any degradation of the organics molecules. For the light exposure, an optical fiber was brought close to the nanogap electrodes inside the glovebox. For the blue light irradiation, we focused the light from a xenon lamp (150 W) to the optical fiber, and we used a dichroic filter centered at 480 nm with a bandwidth of 10 nm (ref 480FS10-50 from LOT Oriel). For UV light irradiation, the optical fiber was coupled with a power LED (ref M365F1 from Thorlabs). This LED has a wavelength centered at 365 nm and a bandwidth of 7.5 nm. At the output of the optical filter, the NPSANs were irradiated on about 1 cm$^2$ at power density of ~0.1 mW/cm² at 480 nm and ~6 mW/cm$^2$ at 365 nm. For the current-voltage (I-V) and reconfigurable logic function experiments, the electrical characteristics were measured with an Agilent 4156C semiconductor parameter analyzer. For HHG measurements, two different signal sources were used : a Tabor waveform generator WW5062 and an Agilent 33220A waveform generator. The output currents were amplified by a low-noise current-voltage amplifier SR570 (Stanford Research Systems) and then analyzed by an Agilent 35670A dynamic signal analyzer. We analyzed the spectra in the 0-100 Hz bandwidth, with a resolution of 1600 points (62.5 mHz/point). The analyzed spectra result from the averaging of 10 spectra in order to increase the signal-to-noise ratio.

**Supporting Information**
Supporting Information is included in this file.

**Acknowledgements.**
We acknowledge financial supports from the EU : FP7-FET-OPEN project SYMONE (# 318597) and H2020 FET-OPEN project RECORD-IT (# GA 664786). We thank the French National Nanofabrication Network RENATECH. We thank Philippe Blanchard (MOLTECH-Anjou, CNRS, Univ. Angers) for the synthesis of AzBT molecules.

# Supporting Information

**Light-stimulable molecules/nanoparticles networks for switchable logical functions and reservoir computing.**

*Y. Viero,[1] D. Guérin,[1] A. Vladyka,[2] F. Alibart,[1] S. Lenfant,[1]*
*M. Calame[2] & D. Vuillaume[1,]\**

1) Institute of Electronics, Microelectronics and Nanotechnology (IEMN), CNRS, Univ. of Lille, Av. Poincaré, 59652 cedex, Villeneuve d'Ascq, France.
\* Corresponding author : dominique.vuillaume@iemn.fr

2) Department of Physics, University of Basel, Klingelbergstrasse 82, 4056 Basel, Switzerland.
and
Transport at Nanoscale Interfaces Laboratory, Empa, Swiss Federal Laboratories for Material Science and Technology, Überlandstrasse 129, 8600 Dübendorf, Switzerland.

**DEVICE FABRICATION**

***NPSANs with metal electrodes.*** First, a ⟨100⟩ oriented silicon wafer was covered with a thermally grown, 200 nm thick, silicon dioxide, formed at 1100 °C during 135 min in a dry oxygen flow (2 L/min) and followed by a postoxidation annealing at 900 °C during 30 min under a nitrogen flow (2 L/min) in order to reduce the presence of defects into the oxide. Second, the e-beam lithography has been optimized by using a 45 nm-thick PMMA (4% 950 K, diluted with anisole with a 5:3 ratio), with an acceleration voltage of 100 keV and an optimized electron beam dose of 370 µC/cm$^2$ for the writing. After the conventional resist development (MIBK:IPA 1:3 during 1 min and rinsed with IPA), a metallic layer (1 nm of titanium and 10 nm of gold) were deposited by e-beam evaporation. Finally after the lift-off process using remover SVCTM14 during 5 h at 80 °C, well-defined coplanar electrodes arranged around a ring with a diameter between 80 to 120 nm were realized (Fig. S1).

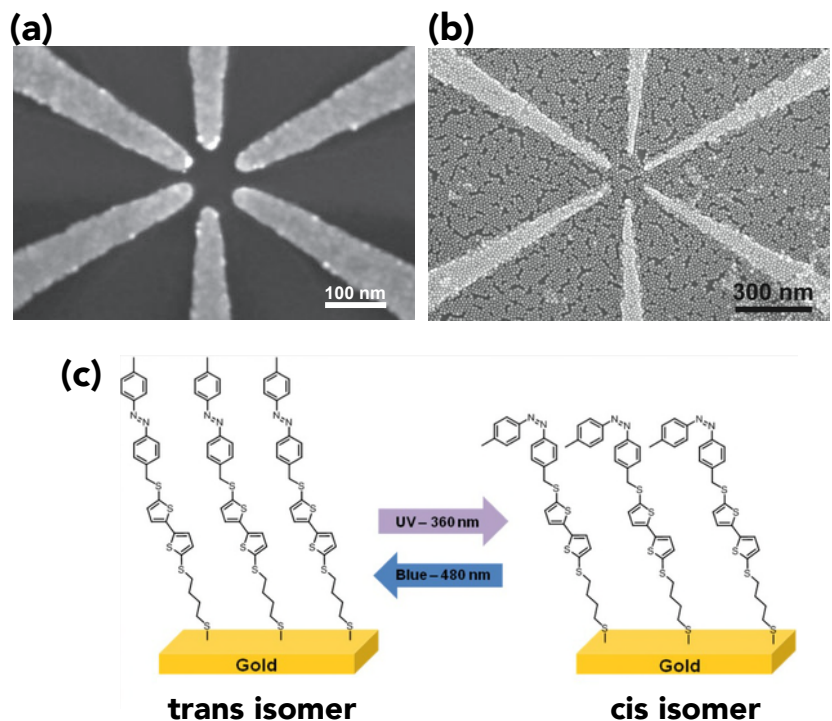

***Fig. S1.*** *SEM images of **(a)** the naked Ti/Au electrodes arranged around a ring of 80 nm in diameter; **(b)** AzBT-covered Au NPs deposited on Ti/Au electrodes (ring of 120 nm). (c) Sketch of the AzBT molecules grafted on the facet of gold NP in the trans and cis isomer.*

## NPSANs with graphene electrodes

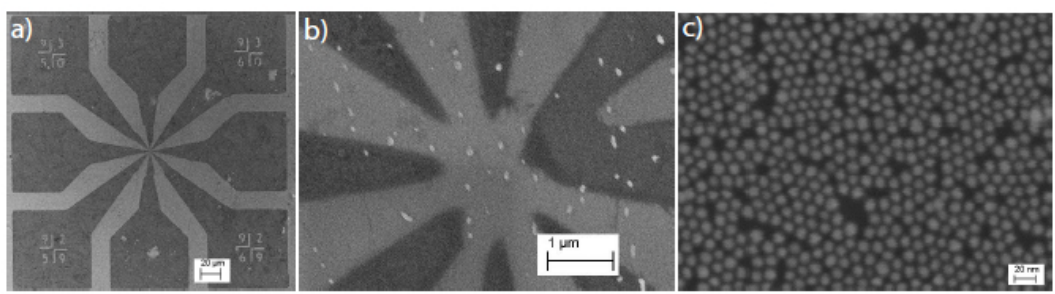

***Fig. S2***. *(a) SEM image of patterned graphene electrodes for 8-terminal hybrid graphene/NPSAN device. **(b)** SEM image of the inner part of the device before NP transfer. Dark areas represent graphene electrodes. **(c)** SEM image of AzBT-covered NP array.*

**ELECTRICAL CHARACTERIZATION.**

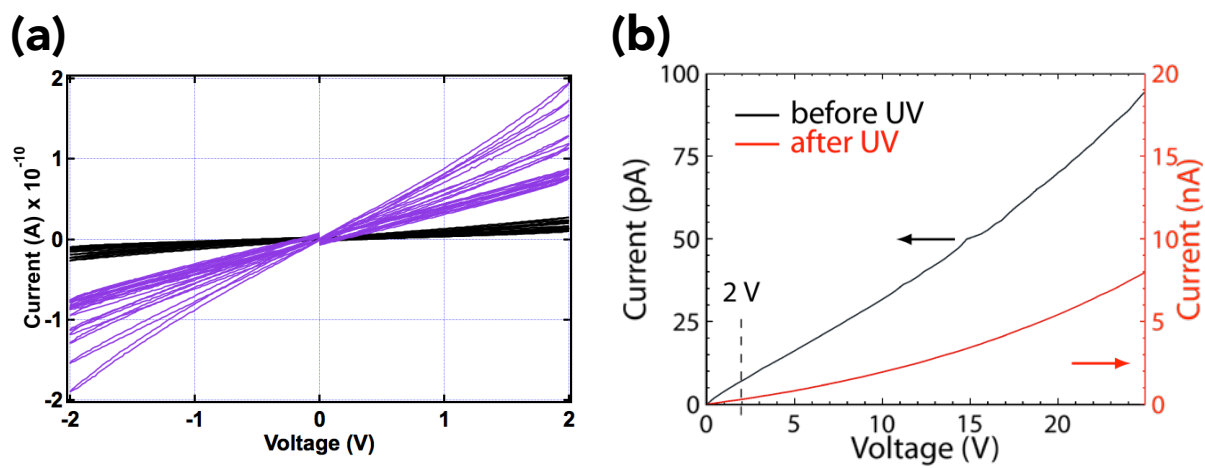

***Fig. S3***. *(a) Current-voltage (I-V) curves measured between all the 15 2-electrode combinations in the NPSANs (Au electrodes) with the AzBT in the trans (black) and cis (purple) isomers. (b) typical I-V curves before (trans AzBT) and after UV light irradiation (cis AzBT) for graphene contacted NPSAN. Since the graphene electrodes are more spaced with a central ring having a diameter of 1 μm (see Fig. S2), the applied voltages is higher than for the NPSANs with Au electrodes.*

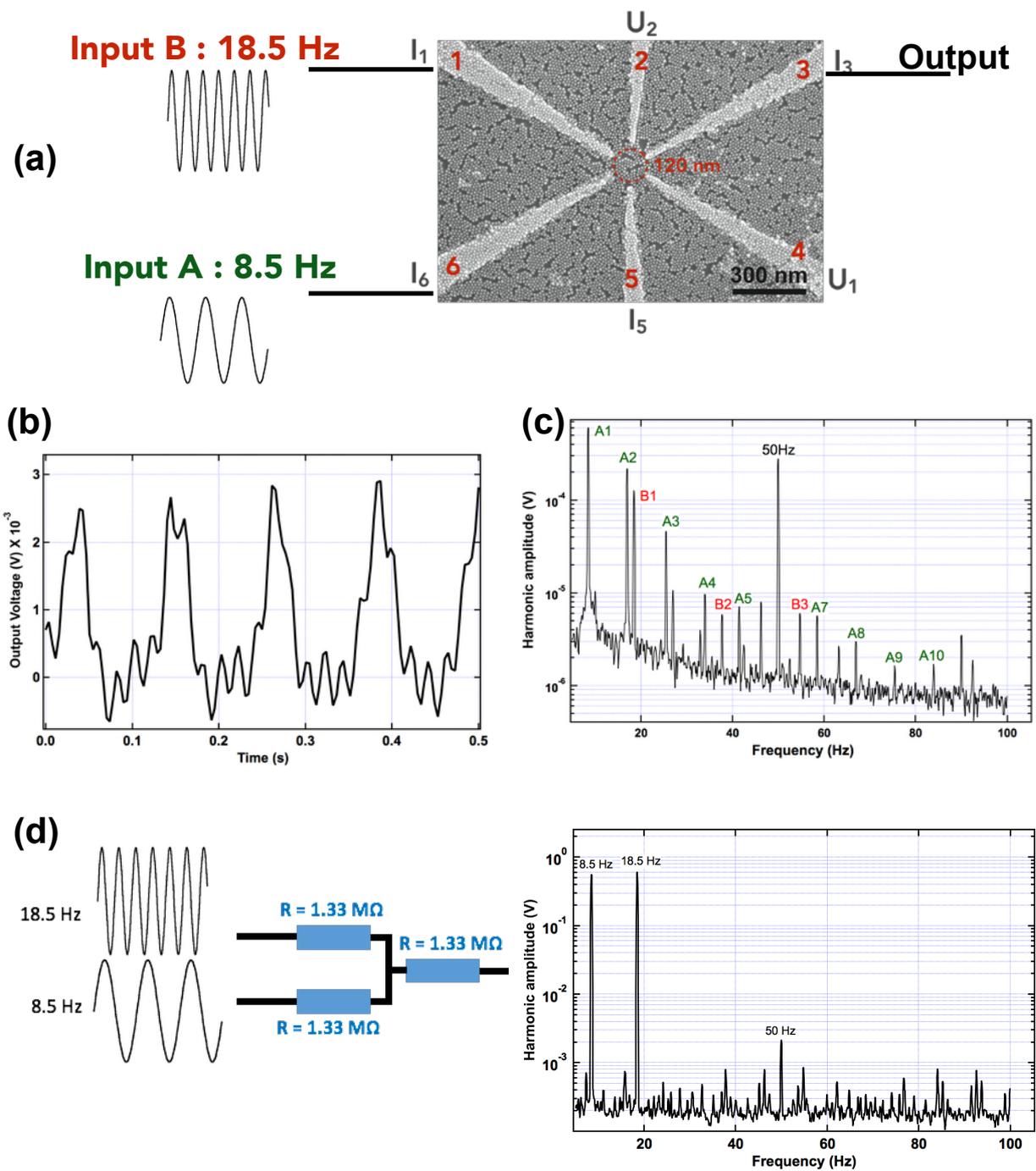

*Fig. S4. (a)* Typical configuration for HHG experiments. Two sinusoidal signals, signal A at 8.5 Hz and signal B at 18.5 Hz (peak-to-peak amplitude $V_{PP}$=8V for both) are applied at electrodes 1 and 6, respectively. Output signal is measured

*on electrode 3. **(b)** Typical output signal. **(c)** HHG spectrum acquired (FFT transform) with the dynamic signal analyzer at electrode 3. The AzBT molecules in the NPSAN are in their trans isomer. The HHG peaks are labeled as Ai (i=1 for the fundamental, i=2 for the 2nd harmonic, etc...) and Bi for harmonics corresponding to the A and B input signals, respectively. **(d)** Control experiment with calibrated resistors.*

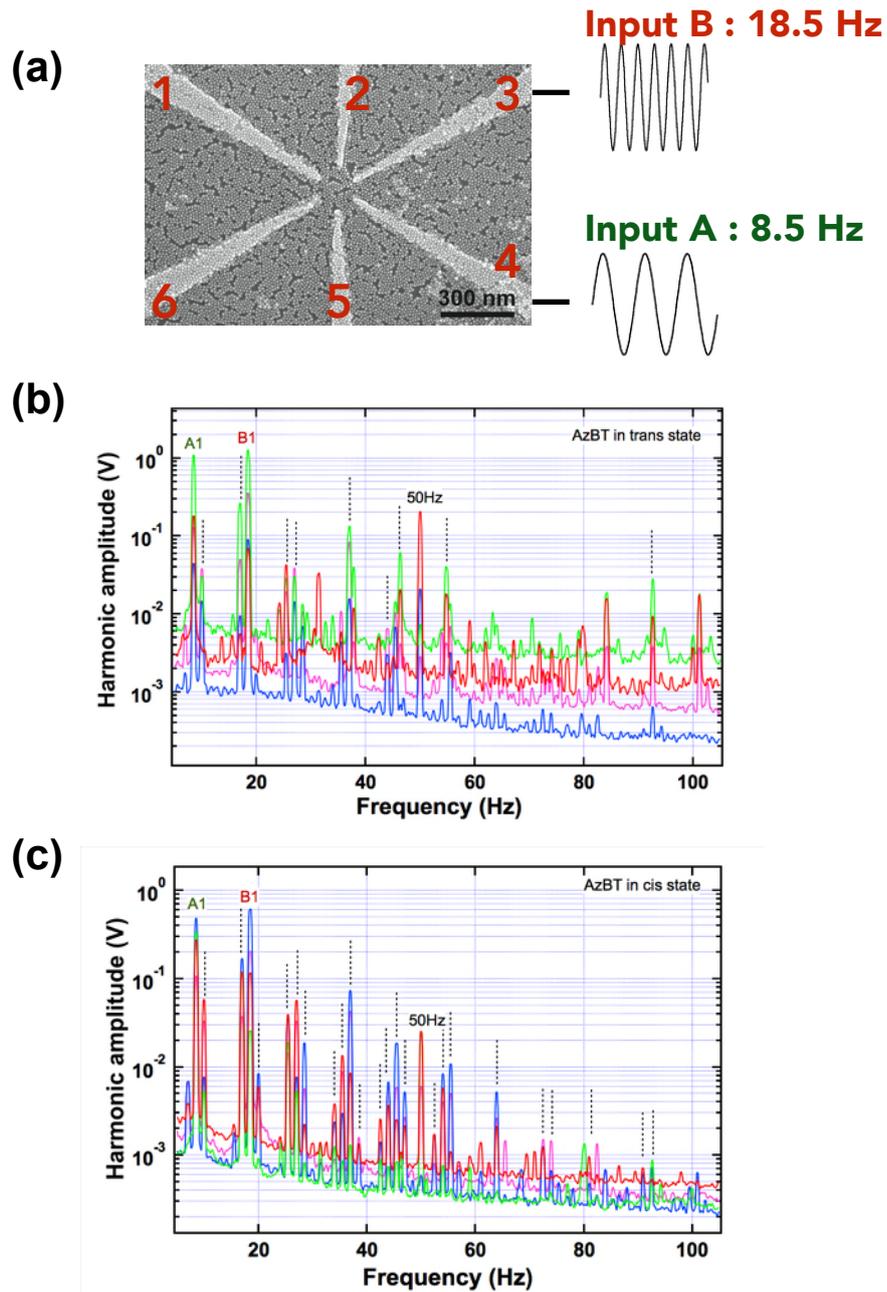

***Fig. S5. (a)*** *Experimental input/output configuration. Comparison of HHG with the AzBT in trans **(b)** and cis isomers **(c)** for the 4 ouputs of the same NPSAN shown in Fig. S5-a. More peaks are observed for the cis isomer, as marked by the vertical dashed lines showing peaks systematically for the 4 traces (outputs) in each cases (trans and cis).*

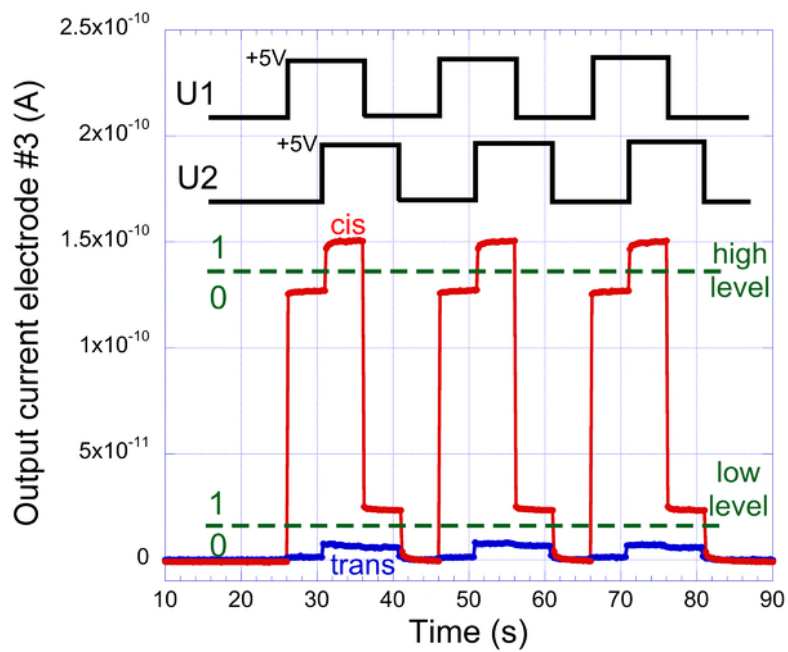

| U1 | U2 | output "trans" | output "cis" high threshold | output "cis" low threshold |
|---|---|---|---|---|
| 0 | 0 | 0 | 0 | 0 |
| 1 | 0 | 0 | 0 | 1 |
| 1 | 1 | 1 | 1 | 1 |
| 0 | 1 | 1 | 0 | 1 |
|   |   | Follower O=U2 | AND O=U1∧U2 | OR O=U1∨U2 |

**Fig. S6** : Analogy of the observed functions in the AzBT-NPSANs with Boolean logic gate. For the output signa with the AzBT in cis, the dashed lines show the threshold defining the "0" and "1" logic levels.

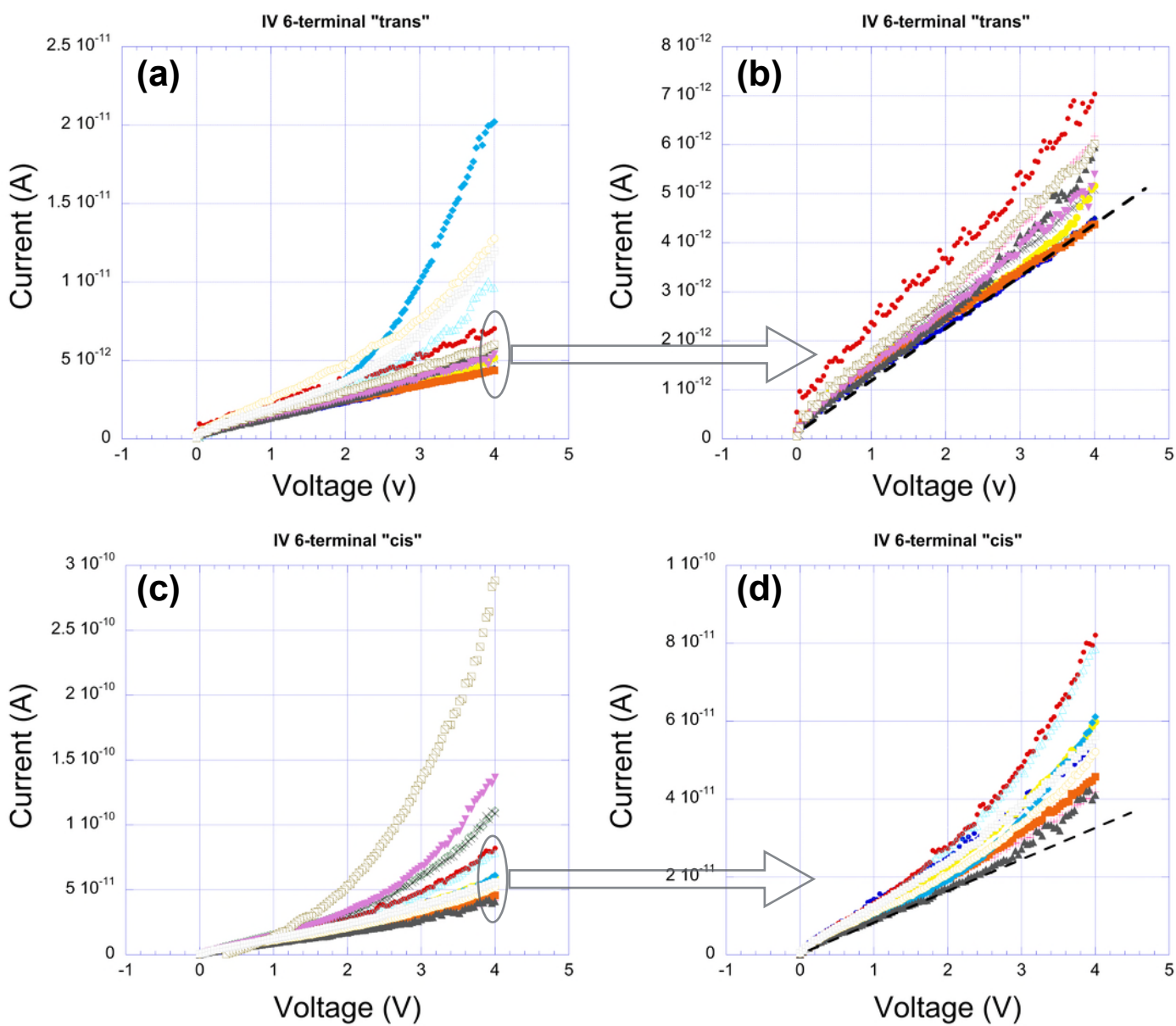

**Fig. S7.** Comparison of the linearity/non-linearity of the 15 possible 2-electrode I-V curves measured on the same NPSAN: **(a-b)** AzBT in *trans*, **(c-d)** AzBt in *cis*.

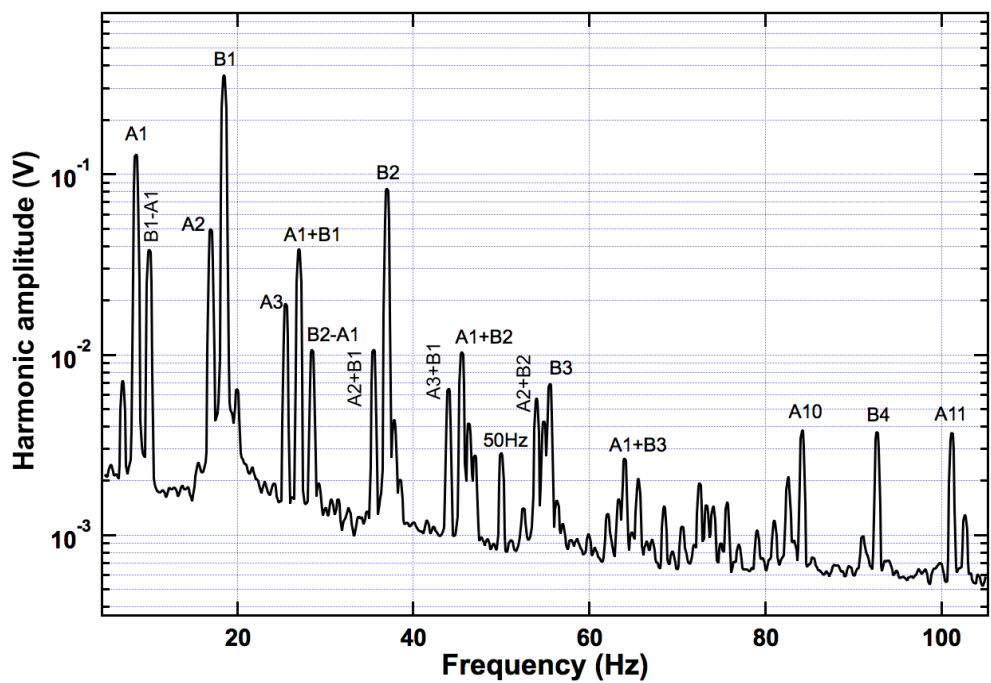

**Fig. S8.** Exemple of IMD for *trans* AzBT-NPSAN, showing an intermodulation distortion product O=4 (e.g. A3+B1, or A1+B3).